\documentclass[pre,aps,twocolumn,groupedaddress,showpacs]{revtex4}

\usepackage{graphicx}
\usepackage{dcolumn}
\usepackage{bm}
\usepackage{amssymb}

\begin{document}

\title{Phase Diagram of Vertically Shaken Granular Matter}

\author{Peter Eshuis,$^1$ Ko van der Weele,$^2$ Devaraj van der Meer,$^1$ Robert Bos,$^1$ and Detlef Lohse$^1$}

\affiliation{\small $^1$ Physics of Fluids, University of
Twente, P.O. Box 217, 7500 AE Enschede, The Netherlands\\
$^2$Mathematics Department, Division of Applied Analysis,
University of Patras, 26500 Patras, Greece}


\date{\today}


\begin{abstract}
A shallow, vertically shaken granular bed in a quasi 2-D container
is studied experimentally yielding a wider variety of phenomena
than in any previous study: (1) bouncing bed, (2) undulations, (3)
granular Leidenfrost effect, (4) convection rolls, and (5)
granular gas. These phenomena and the transitions between them are
characterized by dimensionless control parameters and combined in
a full experimental phase diagram.
\end{abstract}

\maketitle

\newpage

\section{Introduction}

Vertically shaken granular matter exhibits a wealth of fluid-like
phenomena such as undulations~\cite{douady89,clement98,sano05},
wave patterns~\cite{melo94,moon01}, granular Leidenfrost
effect~\cite{eshuis05} and convection rolls~\cite{paolotti04}.
However, while in normal fluids and gases these phenomena are
fully understood, this is much less the case for their granular
counterparts. In order to get a better understanding of the
underlying physics, here we present an experimental overview of
the various effects observed in a vibrated bed of glass beads,
identifying the dimensionless control parameters that govern them.
The main goal of the paper is to construct an experimental phase
diagram in which all the observed phenomena are combined.

\begin{figure}[h!]
  \centering
  \includegraphics[width=6.0cm]{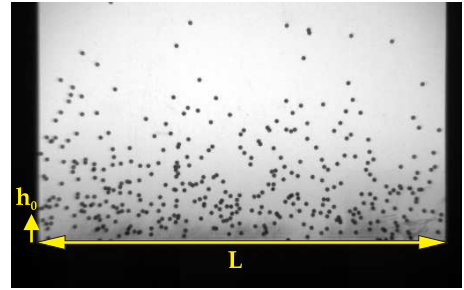}
  \caption{\small The experimental setup in which glass beads
  of diameter $d=1.0~$mm are vibrofluidized. The length of the
  container is $L=101~$mm; the bed height at rest ($h_0$) is
  varied in our experiments such that the aspect ratio $L/h_0$ always
  remains large. The container depth is only
  five particle diameters, making the setup quasi
  two-dimensional.}\label{Intro_Container}
\end{figure}

Our experimental setup (Fig.~\ref{Intro_Container}) consists of a
quasi 2-D perspex container of dimensions $L \times D \times H =
101 \times 5 \times 150~$mm (with $L$ the container length, $D$
the depth, and $H$ the height), partially filled with glass beads
of diameter $d=1.0~$mm, density $\rho=2600~$kg/m$^3$, and
coefficient of normal restitution $e\approx0.95$. The setup is
mounted on a sinusoidally vibrating shaker with tunable frequency
$f$ and amplitude $a$. Most of the experiments presented in this
paper are performed by upsweep-experiments in which the frequency
is linearly increased at $75~$Hz/min. These experiments are
recorded with a high-speed camera capturing $2000$ frames per run;
adequate recording times ($4$ to $16$ seconds) are obtained by
adjusting the frame rate.

The natural dimensionless control parameters to analyze the
experiments are (i) the shaking parameter, or Froude number, $a^2
\omega^2/g \ell$ (with $\omega=2\pi f$ and $g=9.81~$m/s$^2$),
being the ratio of the kinetic energy inserted into the system by
the vibrating bottom and the potential energy associated with a
typical displacement of the particles $\ell$, (ii) the number of
bead layers $F$, (iii) the inelasticity parameter $\varepsilon =
(1-e^2)$, and (iv) the aspect ratio $L/h_0$, where $h_0$ denotes
the bed height at rest. The parameter $\varepsilon$ is taken to be
constant in this paper, since we ignore the velocity dependence
and use the same beads throughout. The aspect ratio varies by
changing the bed height $h_0$ (i.e., the number of layers $F$) but
remains large in all experiments, $L/h_0\gg1$. We will
systematically vary the first two dimensionless parameters, by
changing the amplitude $a$, the frequency $f$, and the number of
layers $F$.

The most intriguing of the four parameters above is the first one,
the shaking parameter, since the typical displacement of the
particles $\ell$ is influenced in a non-trivial way by the
vibration intensity and the number of particle layers. For
\emph{mild} fluidization the displacement of the particles is
determined by the amplitude of shaking $a$, since the bed closely
follows the motion of the bottom. The energy ratio in this case
becomes identical to the well known dimensionless shaking
acceleration:
\begin{equation}\label{eq_Gamma}
\Gamma=\frac{a\omega^2}{g}.
\end{equation}
For \emph{strong} fluidization the particles no longer follow the
bottom, so (instead of $a$) some intrinsic length scale needs to
be taken for $\ell$, such as the particle diameter $d$. This leads
to the dimensionless shaking strength $S$~\cite{eshuis05}:
\begin{equation}\label{eq_S}
S=\frac{a^2\omega^2}{gd}.
\end{equation}
At intermediate fluidization, we will encounter phenomena in which
there is a competition of length scales. In this region the
transitions are affected by changing one of the competing length
scales, meaning that the choice of the appropriate shaking
parameter is not a priori clear. This will become an issue in
particular for the transition from undulations to the granular
Leidenfrost effect described in Section~\ref{sec_Leidenfrost}.

In the following Sections, the various phenomena observed in our
system are discussed one by one, in the order in which they appear
as the fluidization is increased: bouncing bed
(Section~\ref{sec_Bouncing}), undulations (\ref{sec_Undulations}),
granular Leidenfrost effect (\ref{sec_Leidenfrost}), convection
rolls (\ref{sec_Convection}), and granular gas (\ref{sec_Gas}).
Finally, in Section~\ref{sec_Phasediagram} all five phenomena will
be combined in a phase diagram of the relevant shaking parameter
versus the number of layers.

\section{Bouncing Bed}\label{sec_Bouncing}

\begin{figure}[b!]
  \centering
  \includegraphics[width=6.0cm]{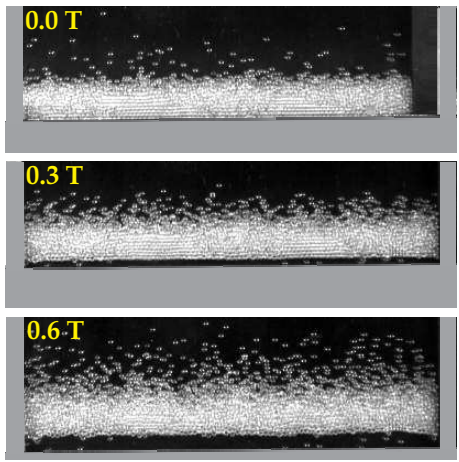}
  \caption{\small Time-series of a bouncing bed for $F=8.1$ layers
  of $d=1.0~$mm glass beads
  at shaking acceleration $\Gamma=2.3$ ($a=4.0~$mm, $f=12.0~$Hz).
  The phase of the sinusoidally vibrating bottom is indicated in each snapshot, where
  $T$ is the period of shaking ($y_{bottom}(t)=a\sin(2\pi t/T)$).
  The friction between the particles and the container walls causes the
  downward curvature of the bed close to the sidewalls that is visible in the lower snapshot.
  [Enhanced online: link to movie of the bouncing bed.]}\label{Bouncing_Example}
\end{figure}

For shaking accelerations $\Gamma \leq 1$ (and even for $\Gamma$
slightly above $1$) the granular bed behaves as a solid, co-moving
with the vibrating bottom and never detaching from it. In order to
detach, the bottom must at some point during the cycle have a
downward acceleration that overcomes gravity (as for a single
bouncing
ball~\cite{holmes82,mehta90,luck93,warr95,warr96,falcon98,geminard03,giusepponi05,zehui06})
\emph{plus} the friction between the bed and the walls of the
container. These walls carry a considerable portion of the bed
weight, as described by the Rayleigh-Jansen model~\cite{duran00}.
Once the detachment condition is fulfilled, the bed bounces in a
similar way as a single particle would do: We call this a bouncing
bed, see Fig.~\ref{Bouncing_Example}.

\begin{figure}[t!]
  \centering
  \includegraphics[width=7.0cm]{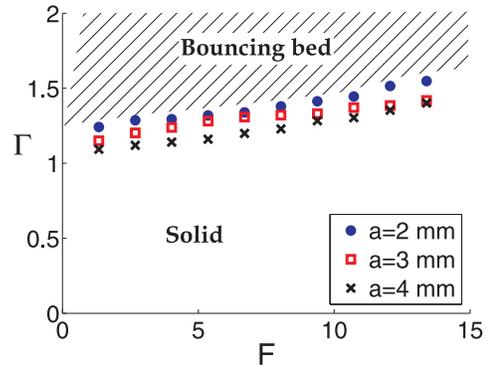}
  \caption{\small (color online). The transition from solid behavior to
  bouncing bed is governed by the shaking parameter $\Gamma$.
  The critical value (here determined for three fixed amplitudes
  $a=2.0 ,3.0 ,4.0~$mm) increases roughly linearly with the number
  of particle layers $F$.}\label{Bouncing_Onset_Gamma}
\end{figure}

The value of $\Gamma$ at which the transition from solid to
bouncing bed occurs has been determined by gradually increasing
the frequency $f$ (for three fixed shaking amplitudes $a=2.0,~3.0$
and $4.0~$mm). The onset value grows with the number of layers
$F$, as shown in Fig.~\ref{Bouncing_Onset_Gamma}. The reason for
this is the larger contact area with the front- and sidewalls
causing a proportionally higher frictional force. Indeed, the
onset value of $\Gamma$ is seen to increase roughly linearly with
the number of layers.

Figure~\ref{Bouncing_Onset_Gamma} indicates that for the current
transition (which occurs at mild fluidization) $\Gamma$ is a good
dimensionless parameter, as explained in the Introduction. It is
not ideal, as exemplified by the fact that the onset values do not
exactly coincide for the different amplitudes of shaking, but for
a different choice of the shaking parameter ($S$) the onset values
differ much more.

\section{Undulations}\label{sec_Undulations}

Starting from a bouncing bed and increasing the shaking frequency
$f$, three different phenomena are observed: (a) For $F\leqslant3$
layers the bed is vaporized and becomes a granular gas
(Section~\ref{sec_Gas}), (b) for $3<F\leqslant6$ convection rolls
form (Section~\ref{sec_Convection}), and (c) for $F>6$ layers the
bed develops undulations (also called "arching" or "ripples" in
the
literature~\cite{douady89,thomas89,pak93,clement96,luding96,wassgren96,aoki96,clement98,hsiau98a,kim99,sano99,hill00,clement00,moon01,ugawa02,ugawa03,deng03,jung04,kanai05,sano05}),
which we will cover in this section.

\begin{figure}[t!]
  \centering
  \includegraphics[width=6.0cm]{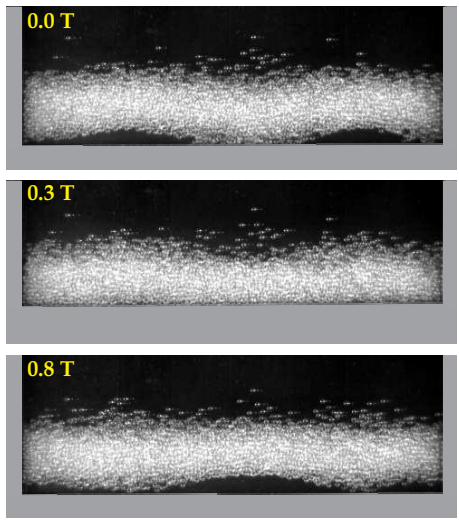}
  \caption{\small One complete standing-wave cycle of the $n=4$ undulation mode for $F=9.4$
  particle layers at $\Gamma=12$ ($a=2.0~$mm, $f=39.3~$Hz).
  The undulation cycle takes $2/f$, i.e., twice the period of shaking.
  [Enhanced online: link to movie of the $n=4$ undulation mode.]}\label{Und_Example}
\end{figure}

In the undulations regime, the granular bed shows standing wave
patterns similar to a vibrating string as shown in
Fig.~\ref{Und_Example}. The standing waves oscillate at twice the
period of shaking, and are therefore also known as
$f/2$-waves~\cite{melo94,moon01}. The container (length $L$)
accommodates an integer number $n$ of half wavelengths of the
granular string:
\begin{equation}
L=n\frac{\lambda}{2},~~~~n=1,2,3,\ldots
\end{equation}
where $\lambda$ is the length of one arch in the undulation
pattern. This $\lambda$ represents a new length scale in the
system besides the shaking amplitude $a$ and the particle diameter
$d$. Unlike these previous length scales, $\lambda$ is connected
to the elastic properties of the particles, which play an
important role in the undulations.

\begin{figure}[t!]
  \centering
  \includegraphics[width=8.5cm]{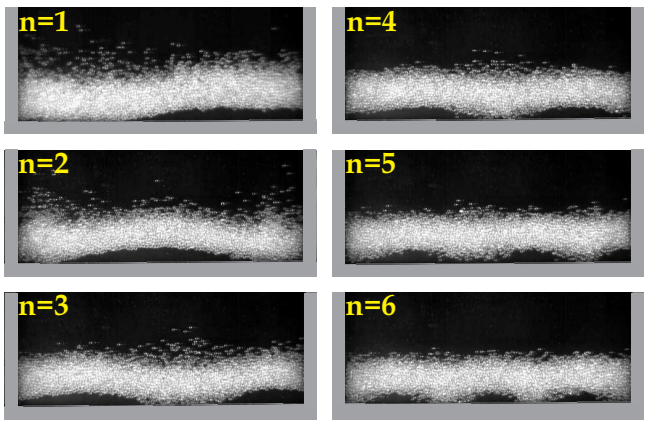}
  \caption{\small Six successive undulation modes, for $F=9.4$ layers
  and $a=2.0~$mm, at shaking frequencies $f=29.0$, $32.6$, $38.2$, $39.3$, $46.1$, $50.2~$Hz.
  The mode number $n$ (the number
  of half-wavelengths fitting the container length $L$)
  increases with the shaking intensity.}\label{Und_Modes_Examples}
\end{figure}

Generally the first undulation to be formed is the $n=1$ mode, and
for increasing fluidization the higher modes depicted in
Fig.~\ref{Und_Modes_Examples} successively appear. They are
triggered by the horizontal dilatancy the bed experiences when it
collides with the vibrating bottom~\cite{sano05}: the string of
particles along the bottom \emph{dilates} and is forced to form an
arch. Using this physical picture, Sano~\cite{sano05} was able to
derive a theoretical form of the undulation modes, which
qualitatively agrees with the form of the experimental ones in
Fig.~\ref{Und_Modes_Examples}. We observe that each collision with
the bottom causes a shock wave through the bed at a roughly
constant speed of $v=\lambda f=2~$m/s. This sends compaction waves
along the arch, starting out from the lower parts and meeting in
the center. At this point the waves bring each other to a halt and
the center falls down to the bottom. (At the same time, the
previous lower parts are now elevated.) This occurs after one
shaking period and the collision with the bottom generates new
shock waves, repeating the series of events. It takes precisely
two periods of shaking to complete one full oscillation of the
undulation pattern.

\begin{figure}[t!]
  \centering
  \includegraphics[width=7.0cm]{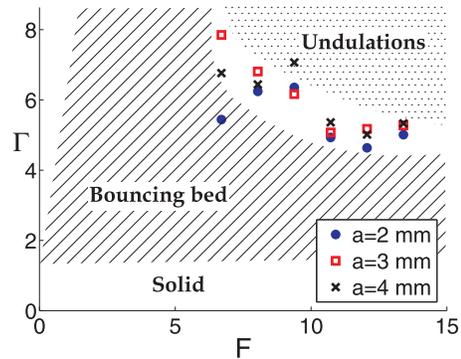}
  \caption{\small (color online).
  The transition from bouncing bed to undulations in the
  $(\Gamma,F)$-plane, for three fixed values of the shaking amplitude ($a=2.0, 3.0, 4.0~$mm).
  The critical value of the shaking acceleration $\Gamma$
  decreases with growing number of particle layers $F$, since the horizontal dilation of the bottom
  layer (required to trigger undulations, see text) becomes more pronounced
  as a result of the extra
  layers on top. }\label{Und_Onset}
\end{figure}

\begin{figure}[t!]
  \centering
  \includegraphics[width=7.0cm]{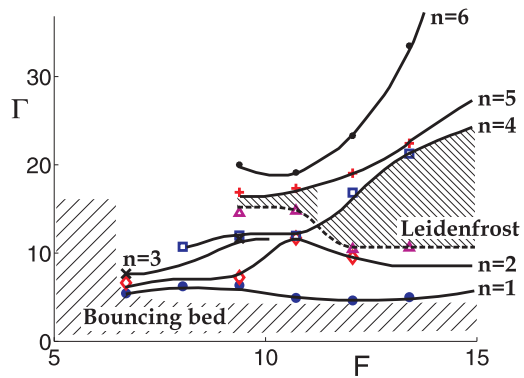}
  \caption{\small (color online). The onset of the successive undulation modes
  $n=1,2,..,6$ at a fixed shaking amplitude $a=2.0~$mm. The mode number $n$
  increases with the shaking parameter $\Gamma$, but occasionally the
  undulations give way to the granular Leidenfrost effect (the hatched regions
  above the dashed curve), where a dense cluster without any arches is
  floating on a uniformly dilute granular layer.}\label{Und_Onset_AllModes}
\end{figure}

First we focus on the transition from the bouncing bed behavior to
the undulations, i.e., on the appearance of the $n=1$ mode. In
Fig.~\ref{Und_Onset} this transition is shown in the
$(\Gamma,F)$-phase diagram for three fixed amplitudes of shaking,
$a=2.0$, $3.0$ and $4.0$~mm. We observe that the onset value of
$\Gamma$ decreases with growing number of layers $F$. The reason
for this is that the necessary horizontal dilation (of the lower
layer) upon impact with the bottom is more readily accomplished
due to pressure from the extra layers on top.

It is seen in Fig.~\ref{Und_Onset} that the data for the three
different shaking amplitudes coincide reasonably, except at the
threshold value of $F=6$ layers. Presumably, at this small value
of $F$ the dilation can only become sufficient if the density is
locally enhanced by a statistical fluctuation; when the experiment
would be repeated many times the agreement between the averaged
data for various $a$ is expected to become better. For $F<6$
layers no undulations are found, since the particle density is
then definitely too small (even in the presence of fluctuations)
to reach the required level of dilation.

Now we come to the higher undulation modes.
Figure~\ref{Und_Onset_AllModes} displays the observed modes for
shaking amplitude $a=2.0~$mm. As already observed in
Fig.~\ref{Und_Modes_Examples} the mode number $n$ increases for
growing $\Gamma$. However, the sequence of modes is seen to be
interrupted somewhere in the middle: Here the undulation pattern
gives way to the granular Leidenfrost state~\cite{eshuis05}, in
which a cluster of slow particles is floating on top of a dilute
layer of fast particles. Normally, this state appears at the end
of the undulation regime (see Section~\ref{sec_Leidenfrost}), but
when a certain standing wave pattern is energetically unfavorable
the system chooses the Leidenfrost state instead. In
Fig.~\ref{Und_Onset_AllModes} we see that this happens to the
$n=3$ undulation, which is completely skipped from the sequence
for $F \gtrsim 12$ layers. This can be understood from the fact
that the $n=3$ mode has an antinode at the sidewall (i.e., a
highly mobile region), whereas the friction with the wall tends to
slow down the particles here. This inherent frustration gives rise
to the appearance of the granular Leidenfrost effect.

Likewise, the small Leidenfrost region for $9 \leq F \lesssim 12$
below the onset line of the $n=5$ undulation may well be the
result of a frustrated $n=5$ mode here. The frustration is however
not strong enough to skip the mode as in the $n=3$ case. In our
experiments, we find that the intermediate regions of the
Leidenfrost state become smaller for larger shaking amplitude $a$.
For $a=4.0$~mm they have disappeared altogether from the
undulation regime, as we will show in
Section~\ref{sec_Phasediagram}.

\section{Granular Leidenfrost effect}\label{sec_Leidenfrost}

\begin{figure}[t!]
  \centering
  \includegraphics[width=6.0cm]{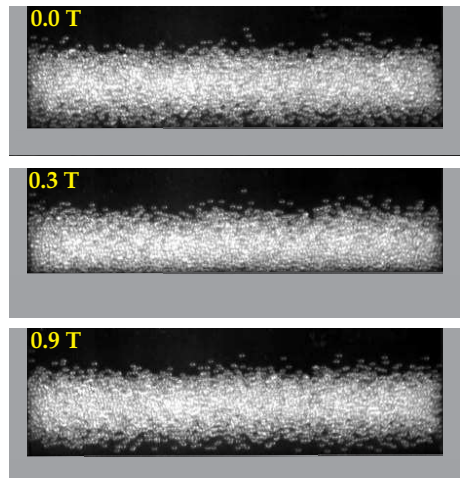}
  \caption{\small Snapshots of the granular Leidenfrost
  effect for $F=8.1$ particle layers shaken at $f=43.0~$Hz and
  $a=3.0~$mm (corresponding to a dimensionless acceleration
  $\Gamma=22$ or shaking strength $S=67$). A dense cluster is elevated and supported by a
  dilute layer of fast particles underneath. The cluster never
  touches the vibrating bottom, which makes this state
  distinctively different from the bouncing bed or the undulations.
  [Enhanced online: link to movie of the granular Leidenfrost effect.]}\label{Leidenfrost_Example}
\end{figure}

\begin{figure}[t!]
  \centering
  \includegraphics[width=6.2cm]{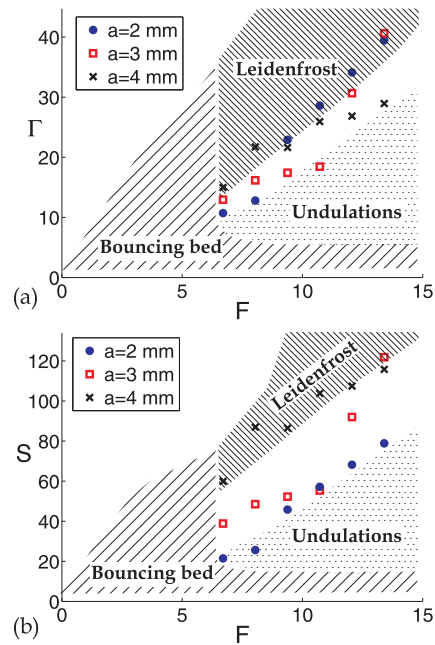}
  \caption{\small (color online). The transition from undulations to
  the granular Leidenfrost effect for increasing frequency $f$ and
  fixed amplitude $a=2.0, 3.0, 4.0~$mm: (a) In the $(\Gamma,F)$-plane,
  (b) in  the $(S,F)$-plane. Since in our experiments the
  Leidenfrost state always originates from the
  undulation regime, the same minimum number of layers is needed:
  $F>6$. The critical values of $\Gamma$ and $S$ increase with $F$,
  since a higher energy input is required to elevate a larger
  cluster.}\label{Leidenfrost_Onset}
\end{figure}

When the shaking frequency is increased beyond a critical level,
the highest undulation mode becomes unstable and we enter the
granular Leidenfrost regime~\cite{eshuis05}: Here a dense cloud of
particles is elevated and supported by a dilute gaseous layer of
fast beads underneath, see Fig.~\ref{Leidenfrost_Example}. The
bottom layer of the undulations is completely evaporated and forms
the gaseous region on which the cluster floats. The phenomenon is
analogous to the original Leidenfrost effect in which a water
droplet hovers over a hot plate on its own vapor layer, when the
temperature of the plate exceeds a critical
value~\cite{leidenfrost66}. The vaporized lower part of the drop
provides a cushion to hover on, and strongly diminishes the heat
contact between the plate and the drop, enabling it to survive for
a relatively long time.

In Fig.~\ref{Leidenfrost_Onset} the transition from the
undulations to the granular Leidenfrost state is shown, both in
the $(\Gamma,F)$ and in the $(S,F)$-plane. Despite the fact that
we have left the mild fluidization regime behind, $\Gamma$ still
appears to be the governing shaking parameter, since the data for
the different amplitudes ($a=2.0, 3.0, 4.0$~mm) collapse better on
a single curve in the $(\Gamma,F)$ than in the $(S,F)$-plane. In
fact, the critical $S$-values in the latter plane show a
systematic increase for growing amplitude $a$.

This is in contrast to the observations on the granular
Leidenfrost effect in a previous study of smaller aspect
ratio~\cite{eshuis05,eshuisPG}, for $d=4.0~$mm glass beads in a
$2$-D container, where the phase transition was shown to be
governed by the dimensionless shaking strength $S$. In that case
the Leidenfrost state was reached directly from the solid,
bouncing bed regime, without the intermediate stage of
undulations. Presumably this was due to the much smaller aspect
ratio $L/h_0$, which was in the order of $1$ (against $L/h_0 \sim
10$ in the present Leidenfrost
experiments)~\cite{footnote_meerson}. Another important difference
was that the depth of the setup was just slightly more than one
particle diameter (against $5$ diameters in the present setup), so
the motion of the granular bed was much more restricted; indeed,
the floating cluster in Ref.~\cite{eshuis05} showed a distinctly
crystalline packing. It may be concluded, as already remarked in
the Introduction, that the Leidenfrost effect lies in the regime
of intermediate fluidization, where both $\Gamma$ and $S$ are
candidates to describe the behavior of the granular bed. The
proper choice of the shaking parameter here depends not only on
the degree of fluidization, but also on the dimensions of the
specific system investigated.

\section{Convection rolls}\label{sec_Convection}

In our experiments, granular convection rolls are formed at high
fluidization from either (a) the bouncing bed (for $3<F\leqslant6$
layers) or (b) the granular Leidenfrost effect (for $F>6$). In
both cases the onset of convection is caused by a set of particles
in the cluster that are more mobile (higher granular temperature)
than the surrounding area, creating an opening in the bed. These
particles have picked up an excess of energy from the vibrating
bottom (due to a statistical fluctuation) and collectively move
upwards, very much like the onset of Rayleigh-B\'{e}nard
convection in a classical fluid heated from
below~\cite{normand77,swift77,bodenschatz00,rogers00,bormann01,oh03,mutabazi06}.
This upward motion of the highly mobile beads must be balanced by
a downward movement of neighboring particles, leading to the
formation of a convection roll.

\begin{figure}[t!]
  \centering
  \includegraphics[width=6.0cm]{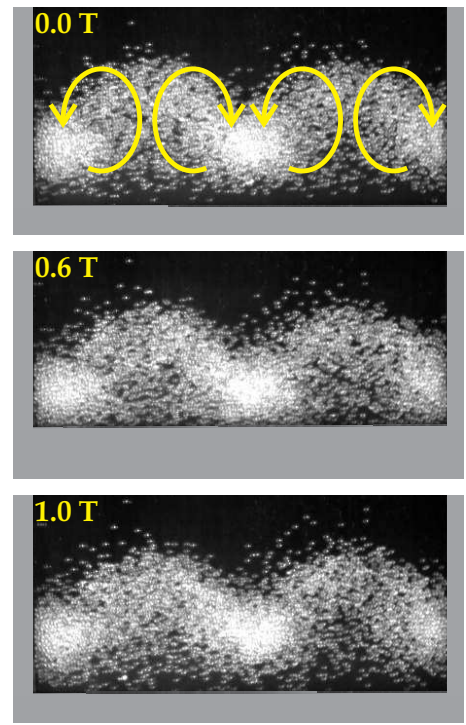}
  \caption{\small Granular convection for $F=8.1$ layers at $f=73.0~$Hz and
  $a=3.0~$mm (dimensionless shaking strength $S=193$), showing
  four counter-rotating rolls. The beads move up in the dilute
  regions (high granular temperature) and are sprayed sideways to
  the three dense clusters (low granular temperature).
  In our system two clusters are always located near the sidewalls, which have
  a relatively low granular temperature due to the extra dissipation.
  [Enhanced online: link to movie of granular convection.]}\label{Convection_Example}
\end{figure}

The downward motion is most easily accomplished at the sidewalls,
due to the extra source of dissipation (i.e., the friction with
the walls), and for this reason the first convection roll is
always initiated near one of the two sidewalls. Within a second,
this first roll triggers the formation of rolls throughout the
entire length of the container, leading to a fully developed
convection pattern as in Fig.~\ref{Convection_Example}.

Granular convection has been studied extensively at mild
fluidization~\cite{clement91,gallas92,taguchi92,knight93,luding94,hayakawa95,ehrichs95,bourzutschky95,aoki96,knight96,lan97,aoki98,bizon98,ramirez00,hsiau00,wildman01,sunthar01,he02,garcimartin02,talbot02,hsiau02,ohtsuki03,khain03,cordero03,miao04,tai04,risso05},
for which the convection is principally boundary-driven. However,
the buoyancy-driven convection observed here occurs at \emph{high}
fluidization and this has been reported much more rarely in the
literature. We are aware of only one numerical study by Paolotti
\emph{et al.}~\cite{paolotti04} and here present the first
experimental observations. In the numerical model by Paolotti
\emph{et al.} the container walls were taken to be perfectly
elastic, leading to convection patterns in which the rolls were
either moving up or down along the sidewalls, whereas in our
system (with dissipative walls) they always move down.

\begin{figure}[t!]
  \centering
  \includegraphics[width=7.0cm]{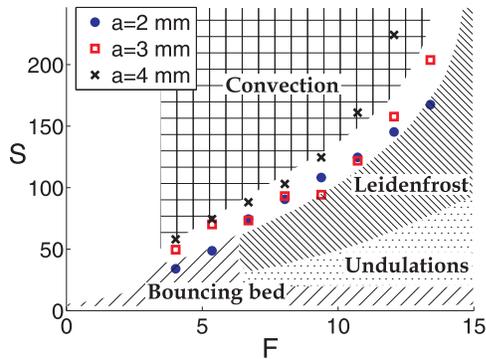}
  \caption{\small (color online). The transition towards granular
  convection from the bouncing bed ($3<F\leqslant6$) and the Leidenfrost
  state ($F>6$) in the $(S,F)$-plane, for fixed shaking amplitude $a=2.0, 3.0$ and $4.0~$mm.
  Just as for the Leidenfrost transition, the convection sets in at higher values
  of $S$ as the number of layers $F$ is increased,
  because a higher dissipation must be overcome for larger bed heights.}\label{Convection_Onset_S}
\end{figure}

Figure~\ref{Convection_Onset_S} shows the transition to convection
in the $(S,F)$-plane, starting from either the bouncing bed or the
Leidenfrost state, which are taken together because the transition
dynamics is the same in both cases. This is the first instance
that the data points (acquired for all shaking amplitudes
$a=2.0,~3.0$ and $4.0~$mm) collapse better for the shaking
parameter $S$ than for the dimensionless acceleration $\Gamma$,
meaning that $S$ is the preferred control parameter for the
convection transition.

The onset values of $S$ grow with the number of layers $F$,
because for large $F$ more energy input from the vibrating bottom
is necessary to break through the larger dissipation in the
granular bed and trigger the first convection roll. Related to
this, the number of rolls in the convection pattern decreases for
growing $F$: Due to the larger total dissipation, the dense
clusters of each roll grow in size. Hence the convection rolls
become wider, meaning that less rolls fit into the container.

\begin{figure}[t!]
  \centering
  \includegraphics[width=8.5cm]{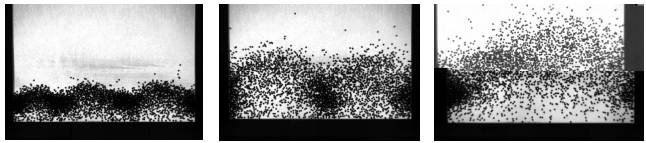}
  \caption{\small Convection patterns for $F=18.8$ layers of $1.0$~mm stainless steel beads at
    three consecutive shaking strengths: $S=58$ ($a=2.0$~mm, $f=60.0$~Hz),
    $S=130$ ($a=3.0$~mm, $f=60.0$~Hz), and
    $S=202$ ($a=4.0$~mm, $f=56.0$~Hz). For increasing $S$ the convection rolls
    expand, hence a smaller number of them fits into the
    container. The steel beads behave qualitatively (but not
    quantitatively) the same as the glass beads used in the rest
    of the paper.}\label{Convection_Roll_Dependence}
\end{figure}

When, for a given number of layers $F$, the shaking strength $S$
is increased (either via the frequency $f$ or the amplitude $a$),
the number of rolls in the convection pattern becomes smaller.
This is illustrated in Fig.~\ref{Convection_Roll_Dependence}: The
higher energy input induces expansion of the convection rolls, and
the number of rolls decreases stepwise as $S$ is increased. The
steps involve two rolls at a time, since the pattern always
contains an even number of rolls due to the downward motion
imposed by the sidewalls.

\section{Granular gas}\label{sec_Gas}

\begin{figure}[t!]
  \centering
  \includegraphics[width=5.0cm]{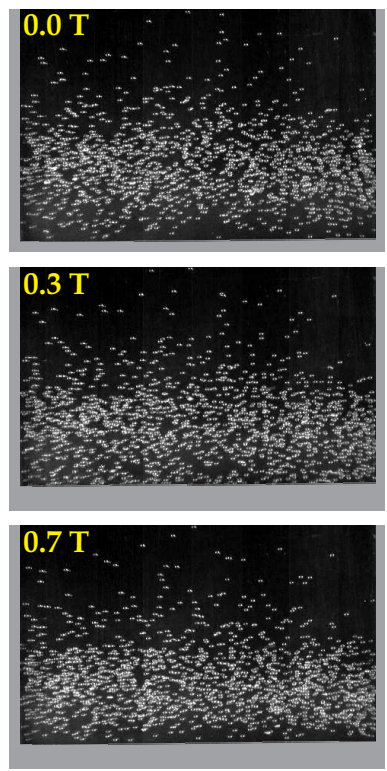}
  \caption{\small Granular gas for $F=2.7$ layers at
  $f=50.0~$Hz and $a=3.0~$mm ($\Gamma=30$), which has originated
  from a bouncing bed by increasing the shaking beyond a critical
  level (see Fig.~\ref{Phase_Diagram_a2a3a4}b).
  With the vibration power available in our system, granular gases
  are only observed for $F\leqslant3$ layers.
  [Enhanced online: link to movie of a granular gas.]}\label{Gas_Example}
\end{figure}

In this section we briefly discuss the fifth and last phenomenon
observed in our system: A granular gas, being a dilute cloud of
particles moving randomly throughout the container as in
Fig.~\ref{Gas_Example}~\cite{goldhirsch93,du95,grossman97,kudrolli97,sela98,eggers99,weele01,meer02,goldhirsch03,hayakawa03}.
This state is only observed for a small number of layers
($F\leqslant3$) and always originates from the bouncing bed
regime. At these small $F$, the bed shows expansion and compaction
during every vibration cycle due to the low total dissipation. At
the critical value of the shaking parameter, the bed expands to
such an extent that it evaporates and forms a gas.

The evaporation of the bouncing bed requires more energy as the
number of layers $F$ increases. The transition seems to be
controlled by the shaking acceleration $\Gamma$ (which also
governs the transition from solid to bouncing bed) rather than the
shaking strength $S$. However, the data points available are too
few ($F\leqslant3$) to make this conclusive. The measurements will
be presented in the full phase diagram of the next section.

\section{Phase diagram}\label{sec_Phasediagram}

\begin{figure}[t!]
  \centering
  \includegraphics[width=7.9cm]{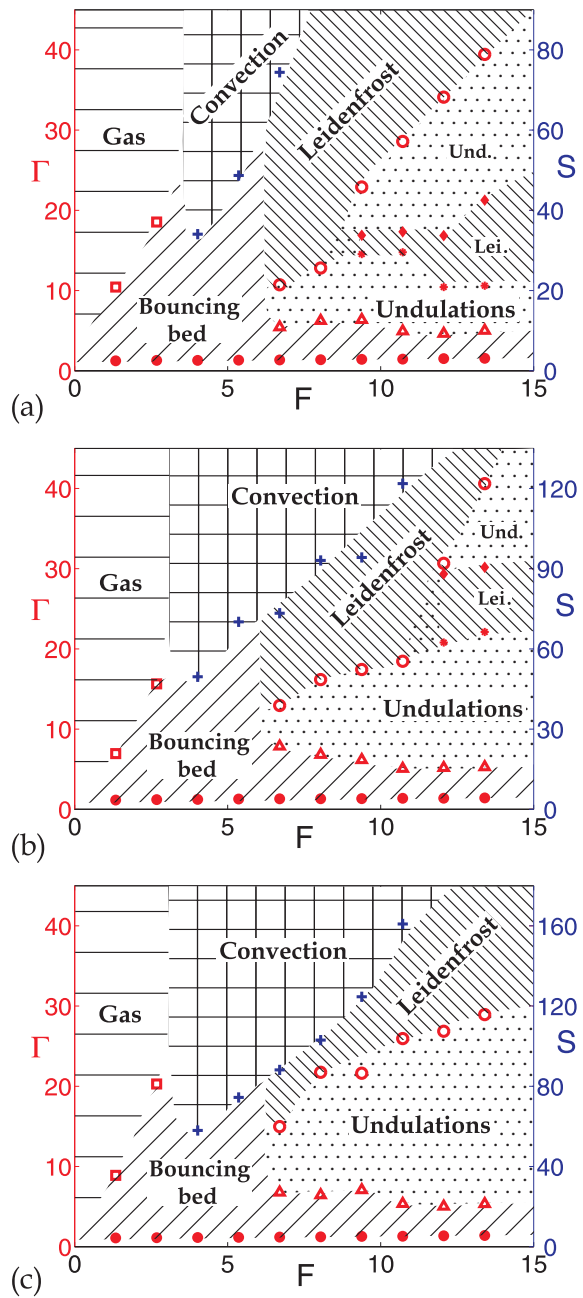}
  \caption{\small (color online). Phase
  diagram of the shallow granular bed at three fixed values of the shaking amplitude: (a) $a=2.0~$mm,
  (b) $a=3.0~$mm, and (c) $a=4.0~$mm. The five phenomena
  studied in this paper are indicated by the different shadings.
  The onset values for bouncing bed, undulations, Leidenfrost
  effect, and gas are governed by the shaking parameter $\Gamma$
  (left vertical axis, red); the onset of convection is
  controlled by $S$ (right vertical axis, blue). The narrow region
  without shading along the horizontal $F$-axis (below the
  bouncing bed regime) corresponds to the solid phase, in which
  the bed never detaches from the vibrating bottom.
  [Enhanced online: link to movie showing all transitions
  of diagram (b), for increasing shaking frequency at $F=8.1$ layers.]}\label{Phase_Diagram_a2a3a4}
\end{figure}

Finally, all the phenomena and associated transitions described in
the previous sections are combined in the phase diagram of
Fig.~\ref{Phase_Diagram_a2a3a4}. Both shaking parameters ($\Gamma$
and $S$) are used in this diagram, each of them indicating the
respective transitions they were found to govern. The parameter
$\Gamma$ is shown along the left vertical axis and the
corresponding data points (the critical $\Gamma$ values) are
colored red. The parameter $S$ is plotted along the right vertical
axis and the corresponding experimental data are colored blue;
this concerns only the $+$-signs at the convection
transition~\cite{footnote_phasediagram}. For comparison the
$\Gamma$-axis is kept the same in all three phase diagrams.

Figure~\ref{Phase_Diagram_a2a3a4} contains three separate phase
diagrams for the three fixed shaking amplitudes we have used
throughout the paper: $a=2.0$, $3.0$, and $4.0~$mm. Most of the
phase transitions are hardly affected, with the exception of the
various transitions between the undulations and the Leidenfrost
state. These transitions lie in the regime of \emph{intermediate}
fluidization, where the system experiences a competition of length
scales: the amplitude $a$, the particle diameter $d$, and
additionally the wavelength of the undulations $\lambda$. This
becomes especially clear in the phase diagram of
Fig.~\ref{Phase_Diagram_a2a3a4}(a) for $a=2.0~$mm where the
competition results in an alternation of states. By increasing $a$
in Fig.~\ref{Phase_Diagram_a2a3a4}(b,c) it becomes the dominant
length scale and the alternation vanishes ultimately.

How does Fig.~\ref{Phase_Diagram_a2a3a4} compare with other phase
diagrams for shaken granular matter in the literature?

First we discuss the experimental phase diagram by Wassgren
\emph{et al.}~\cite{wassgren96} for a bed of $1.28$~mm glass beads
at mild fluidization ($\Gamma\leq10$). For increasing $\Gamma$,
they observe a series of transitions from a solid bed to
undulations ("arching") in qualitative agreement with our own
experiments at mild shaking. Their series of transitions is
interlaced however with several phenomena (Faraday heaping,
surface waves) that are not observed in our system. This is
presumably due to the larger depth of their container ($12.5$
particle diameters, versus $5$ in our container, which means that
their setup deviates considerably from $2$-D) and to the fact that
their bed height was typically an order of magnitude larger than
ours: The lowest aspect ratio $L/h_0$ in their experiments was
$2$, versus $10$ in our system. Hsiau and Pan~\cite{hsiau98}, who
conducted experiments in a similar setup in the mild fluidization
regime, found the same sequence of phenomena as Wassgren \emph{et
al.}~\cite{wassgren96}. Indeed, in three dimensions a much wider
variety of phenomena is observed than in $2$-D
systems~\cite{savage88,melo94,melo95,tsimring97,metcalf97,shinbrot97,bizon98,mujica98,mujica99,aranson99,umbanhowar00,blair00,bruyn01,moon01,kim02,park02,moon04,goldman04,bougie05,wong06}.

Secondly, Sunthar and Kumaran~\cite{sunthar01} construct a phase
diagram (shaking strength vs. number of layers) based on event
driven simulations in a $2$-D system with an aspect ratio
$L/h_0\gtrsim10$, comparable to ours. At low shaking strength,
their phase diagram shows a region where the bed is "homogeneous",
corresponding to the solid and bouncing bed regimes in our
diagram. At higher shaking strength, they find a granular gas for
$F < 5$ and a region of granular convection for $F > 5$. The
gaseous region compares well with the gas region in
Fig.~\ref{Phase_Diagram_a2a3a4}. The convection observed by
Sunthar and Kumaran, however, occurs at a much milder fluidization
than in our system. In contrast to our convection rolls, the
density of their rolls is almost constant, indicating that the bed
behaves more like a fluid than a gas.

Thirdly, Eshuis \emph{et al.}~\cite{eshuis05} construct an
experimental phase diagram (supported by a theoretical model) for
a bed of $4~$mm glass beads in a $2$-D setup. The $(S,F)$-diagram
shows a bouncing solid regime for low shaking strength and a gas
region for small $F$. Between these two phases, for $S\gtrsim16$
and $F>8$, the Leidenfrost regime is located. This is
qualitatively the same as in Fig.~\ref{Phase_Diagram_a2a3a4},
without the regions of undulations and granular convection though.
The fact that these latter phenomena were absent is probably due
to the much smaller aspect ratio ($L/h_0 \thicksim 1$) and the
much stronger confinement to two dimensions, since the depth of
the container was just slightly more than one particle diameter.

Finally, Paolotti \emph{et al.}~\cite{paolotti04} performed a
$2$-D numerical study of a granular bed with aspect ratio
$L/h_0\thickapprox8$, focusing on the transition towards
convection. Their convection rolls show similar arches and
distinct density differences as observed in our experiments.
Starting from strong fluidization, for a given number of layers,
they observe two transitions as the shaking strength is reduced.
First a transition from a non-convective state (presumably a
granular gas) to convection, followed by a transition towards a
non-convective state again, in which the particles remain
localized near the bottom. This latter state is not further
specified, but most probably corresponds to a bouncing bed. In the
phase diagram of Fig.~\ref{Phase_Diagram_a2a3a4} the same sequence
is found if one follows a path from the gas regime to the bouncing
bed via convection.\\

In conclusion, we have constructed the experimental phase diagram
for a vertically shaken shallow granular bed in a quasi 2-D
container, identifying the dimensionless control parameters that
govern the various transitions in this diagram. In the present
work we have concentrated on $\Gamma$ and $S$ (the shaking
parameters), and the parameter $F$ (number of particle layers).
From the discussion above, it may be concluded that also the
aspect ratio is an important control parameter, determining e.g.
the set of different phenomena that a given system is able to
exhibit.

The diagram of Fig.~\ref{Phase_Diagram_a2a3a4} shows the full
range of phases that granular matter can display, behaving either
like a solid, a fluid, or a
gas~\cite{jaeger92,jaeger96,liu98,kadanoff99,aranson06}. A
determination of the dimensionless parameters that govern the
transitions between these phases is a crucial step towards a
better understanding of the physics of vertically shaken granular
matter.

\noindent {\it Acknowledgment:} We thank Stefan Luding for
stimulating discussions. This work is part of the research
program of FOM, which is financially supported by NWO.\\

\newpage


%

\begin{thebibliography}{100}
\bibitem{footnote_phasediagram}
To keep the phase diagram well-proportioned, not all data points
of the Leidenfrost-convection transition ($+$) are presented in
Fig.~\ref{Phase_Diagram_a2a3a4}. The full set of data points can
be seen in Fig.~\ref{Convection_Onset_S}.
\end{thebibliography}


%

\end{document}